\documentclass[11pt]{article}
\usepackage{color}
\usepackage{amssymb}
\usepackage{amsmath}
\usepackage{multirow}
\usepackage{epsfig}
\usepackage{mfpic}
\usepackage{amscd}
\usepackage{feynmf}

\def\Journal#1#2#3#4{{#1} {\bf #2}, #3 (#4)}

\def\etal{{\it et al.}}

\def\APH{\em Annals Phys.}
\def\APN{\em Annalen Phys.}
\def\APJ{\em ApJ.}

\def\AST{\em Astron. J.}

\def\CPL{\em Chin. Phys. Lett.}
\def\CQG{\em Class.Quant.Grav.}

\def\IMD{{\em Int. J. Mod. Phys.} D}

\def\JCA{\em J. Cosmol. Astrop. Phys.}
\def\JHE{\em J. High Ener. Phys.}

\def\JPL{\em JETPhys. Lett.}

\def\NPB{{\em Nucl. Phys.} B}

\def\PLB{{\em Phys. Lett.}  B}
\def\PRC{{\em Phys. Rev.} C}
\def\PRD{{\em Phys. Rev.} D}
\def\PRL{\em Phys. Rev. Lett.}

\def\PRE{\em Phys. Rep.}

\def\be{\begin{equation}}
\def\ee{\end{equation}}
\def\bea{\begin{eqnarray}}
\def\eea{\end{eqnarray}}
\def\bes{\begin{equation*}}
\def\ees{\end{equation*}}
\def\beas{\begin{eqnarray*}}
\def\eeas{\end{eqnarray*}}

\def\tr{\text{tr}}
\def\mg{\mathsf g}
\def\cx{\mathtt X}
\def\ca{\mathtt A}
\def\um{\mathcal U}

\begin{document}
\begin{center}
\Large \bf {Condensation of a Classical Scalar Field After Inflation and Dark 
Energy}\\
\end{center}

\begin{center}
{\it Houri Ziaeepour\\
{Mullard Space Science Laboratory, University College London\\Holmbury St. Mary, Dorking, Surrey
RH5 6NT, UK.\\
Email: {\tt hz@mssl.ucl.ac.uk}}
}
\end{center}


\begin{abstract}
In cosmological context, classical scalar fields are important ingredients for 
inflation models, many candidate models of dark energy, symmetry breaking and 
phase transition epochs, and their consequences such as baryo and 
lepto-genesis. We investigate the formation of these fields by studying the 
production of a light quantum scalar field during the decay of a heavy 
particle. For simplicity it is assumed to be a scalar too. We discuss the 
effects of the decay mode, the thermodynamical state of the decaying field, 
boundary conditions, and related physical parameters on the production and 
evolution of a condensate. For a simplified version of this model we 
calculate the asymptotic behaviour of the condensate and conditions for its 
contribution to the dark energy with an equation of state close to a 
cosmological constant. We also discus the role of the back-reaction from 
interactions with other fields and expansion of the Universe on the evolution 
of the condensate.
\end{abstract}

\begin{fmffile}{fmfgenfydig}
\section{Introduction} \label{sec:intro}
Symmetry breaking, phase transition, and related phenomena such as appearance 
of a dynamical mass (Higgs mechanism), superconductivity, inflation, 
leptogenesis, and quintessence field as a candidate for dark energy in early 
universe and cosmology are based on the existence of a 
{\it classical scalar field}. As the physics of the Universe and its content 
in its most elementary level is quantic, this scalar field, fundamental or 
composite, is always related to a {\it quantum scalar field}. 

A classical field is more than just classical behaviour of a large number of 
scalar particles. In a quantum system, fields/particles are in superposition 
states and are quantum mechanically correlated with each others. Decoherence 
process removes the superposition/correlation between particles. However, 
this does not mean that after decoherence of scalar particles, they will 
behave collectively like a classical field. A simple example is the 
following:

Consider a closed system consisting of a macroscopic amount of unstable 
massive scalar particles that decay to a pair of light scalar particles with 
a global $SU (2)$ symmetry and negligible interaction. If the unstable 
particle is a singlet of this symmetry, the remnant particles are entangled 
by their $SU (2)$ state. After a time much larger than the lifetime of the 
massive particle, the system consists of a relativistic gas of pair entangled 
particles. If a detector measures this $SU (2)$ charge without significant 
modification of their kinetic energy, the entanglement of pairs will break 
i.e. the system decoheres and becomes a relativistic gas. The equation of 
state of a relativistic ideal gas is $w_{rel} = P/\rho \approx 1 / 3$. 
In contrast to an operator in quantum field theory, a classical scalar field 
$\varphi (x)$ is a $C$-number. Its density ${\rho}_{\varphi}$, pressure 
$P_{\varphi}$ and kinetic energy are defined as:
\bea
{\rho}_{\varphi} \equiv K_{\varphi} + V (\varphi) \label{rhophi} \\
P_{\varphi} \equiv K_{\varphi} - V (\varphi) \label{pphi} \\
K_{\varphi} = \frac {1}{2}g^{\mu\nu} \partial_\mu \varphi \partial_\nu \varphi 
\label{kphi}
\eea
$V (\varphi)$ is the potential presenting the self-interaction of the field 
$\varphi (x)$. When it is much smaller than kinetic energy $ K_{\varphi}$ , one 
obtains $P_{\varphi} \approx {\rho}_{\varphi}$. On the other hand, if 
$V (\varphi) \gg K_{\varphi}$, $P_{\varphi} \approx -{\rho}_{\varphi}$. 
Therefore in general, a relativistic gas and a scalar field don't share the 
same equation of state. Thus, the proof of decoherence in a system is not 
enough when a classical scalar field is needed to explain physical phenomena.

According to canonical quantization procedure, classical observables are 
replaced by operators acting on a Hilbert or Fock space of states 
respectively for a single particle and for a multi-particle quantum system. 
The expectation value of these operators are the outcome of measurements. 
Therefore, it is natural to define the classical observable related to a 
quantum scalar field as its expectation value:
\be
\varphi (x) \equiv \langle \Psi|\Phi (x)|\Psi\rangle \label{classphi}
\ee
where $|\Psi\rangle$ is the state of the quantum system - an element of the 
Fock space. In analogy with particles in the ground state in quantum 
mechanics, the classical field $\varphi (x)$ is also called a 
{\it condensate}. Using canonical representation it is easy to see that for 
a free quantum scalar field $\langle \Psi|\Phi|\Psi \rangle = 0$. Therefore, 
a necessary condition for the buildup of a classical scalar field is an 
interaction (see also Appendix \ref{app:a}).

In quintessence models a classical field is the basic content of the model and 
the source of the dark energy. Although in the framework of popular particle 
physics models such as supersymmetry, supergravity, and string theory many 
works have been concentrated on finding candidate scalar field to play the 
role of quintessence~\cite{quincand}, little effort has been devoted to 
understand the necessary conditions for a quantum scalar fields to condense 
in a manner satisfying very special characteristics needed for a quintessence 
field. For instance, such a condensate must have a very small density, much 
smaller than other content of the Universe (smallness problem). Present 
observations show that dark energy has a behaviour close to a cosmological 
constant i.e. with expansion of the Universe either its energy density does 
not change or varies very slowly. Such a behaviour is not trivial. In the 
classical quintessence models usually the potential or other characteristics 
of the models are {\it designed} such that a tracking solution is obtained. 
However, apriori it is not trivial to prove that in the early Universe a 
quantum field could produce a condensate with properties similar to dark 
energy.

The purpose of the present work is to fill the gap between quantum processes 
producing various species of particles/fields in the early Universe - 
presumably during and after reheating - their classical component as defined 
in (\ref{classphi}). In another word we want to see how the microscopic 
properties of matter is related to macro-physics and vis versa. As the 
quantum physics of that epoch is not well known, we consider the simple case 
of a scalar field - quintessence field - in interaction with two other scalar 
fields as a prototype process, and study the evolution of the classical 
component (condensate) of the quintessence field. Between many possible types 
of quantum scalar field and interaction models, we specially concentrate on 
a class of models is which the scalar field is one of the remnants of the 
decay of a heavy particle. Motivations for such a model are the results of 
studying the effects of a decaying dark matter on the equation of states of 
the Universe~\cite{houriustate}. It has been shown that a FLRW cosmology 
with a decaying dark matter and a cosmological constant at late times behaves 
similar to a cosmology with a stable dark matter and a dark energy component 
with $w = P/\rho \lesssim -1$. This is effectively what is concluded at least 
from some of present supernovae observations~\cite{snobs}. Recently, the same 
effect has 
been proved for the general case of interaction between dark matter and dark 
energy~\cite{quindmint}. It has been also shown~\cite{houridmquin} that if a 
decaying dark matter has a small branching factor to a light scalar field, 
this can explain both observed density and equation of state of the dark 
energy without extreme fine-tuning of the potential or coupling constants. 
In other words, such a model solves both the smallness and the coincidence 
problems of the dark energy. These studies however are based on the 
assumption of a classical scalar field (a condensate). The present work 
should complete this investigation by studying the formation and evolution 
of classical component from quantum processes. 

In Sec. \ref{sec:decay} we construct the Lagrangian of a decaying dark matter 
model. We consider two decay modes for the heavy particle and use the closed 
time path integral method to 
calculate the contribution of interactions to the condensate. The same 
methodology has been used for studying inflation models~\cite{infquant}, 
late-time warm inflation~\cite{warminf}, the effects of renormalization and 
initial conditions on the physics of inflation~\cite{infrenorm}, 
baryogenesis~\cite{baryo}, and coarse-grained formulation of 
decoherence~\cite{coarsegrain}. In Sec. \ref{sec:classevol} we obtain an 
analytical expression for the asymptotic behaviour of the condensate and 
discuss the importance of the back-reaction of the quantum state of the 
Universe on the evolution of the condensate and properties of the dark energy. 
We summarize the results in Sec. \ref{sec:conclu}. In Appendix \ref{app:a} 
we make a remark about relation between decoherence and formation of a 
condensate. In Appendixes \ref{app:b} we show that the effect of a non-vacuum 
state on the Green's function can be included in the boundary conditions. 
Appendix \ref{app:c} presents the solution of the evolution equation of the 
field in matter dominated era.

\section{Decay in an Expanding Universe} \label{sec:decay}
We consider a simple decay mode for a heavy particle $X$ with only 2 types of 
particles/fields in the remnants: a light scalar $\Phi$ - light with respect 
to the decaying particle - and another field $A$ of an arbitrary type. In 
fact, in a realistic particle physics model, most probably $A$ will not 
be a final stable state and decays/fragments to other particles. Therefore, it 
should be considered as an intermediate state or a collective notation for 
other fields. In the simplest case studied here all the particles are assumed 
to be scalar. Extension to the case where the decaying particle $X$ and one 
of the remnants are spinors is straightforward. The quintessence field $\Phi$ 
must be a scalar to condensate. Nonetheless, in the extreme densities of 
the Universe after reheating, apriori the formation of Cooper-pair like 
composite scalars is possible if the interaction between spinors is enough 
strong. Thus, $\Phi$ can be such a field, but for the simple model studied 
here we ignore such complexities. 

The simplest decaying modes are the followings:
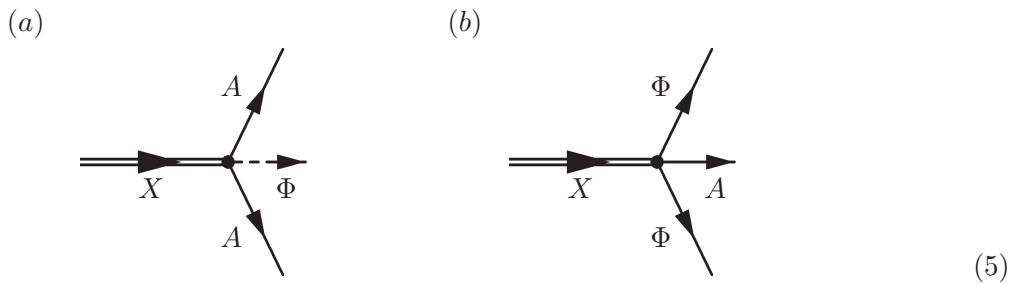
\bea
(a) \quad \quad \quad \quad \quad \quad \quad \quad \quad & \quad \quad \quad 
\quad (b) \nonumber \\
\begin{fmfgraph*}(30,30)
\fmfleft{i1}
\fmfright{o2,o3,o4}
\fmf{dbl_plain_arrow,label=$X$}{i1,v1}
\fmf{fermion,label=$A$}{v1,o2}
\fmf{dashes}{v1,v2}
\fmf{plain_arrow,label=$\Phi$}{v2,o3}
\fmf{fermion,label=$A$}{v1,o4}
\fmfdot{v1}
\end{fmfgraph*} & \quad &
\begin{fmfgraph*}(30,30)
\fmfleft{i1}
\fmfright{o2,o3,o4}
\fmf{dbl_plain_arrow,label=$X$}{i1,v1}
\fmf{plain_arrow,label=$\Phi$}{v1,o2}
\fmf{plain}{v1,v2}
\fmf{fermion,label=$A$}{v2,o3}
\fmf{plain_arrow,label=$\Phi$}{v1,o4}
\fmfdot{v1}
\end{fmfgraph*} \label{decaymode}
\eea
Diagram (\ref{decaymode}-a) is a prototype decay mode when $X$ and 
$\Phi$ shares a conserved quantum number. For instance, one of the favorite 
candidates for $X$ is a sneutrino decaying to a much lighter scalar field 
(e.g. another sneutrino) carrying the same leptonic 
number~\cite{rnu}~\cite{rnu1}. With seesaw mechanism in the superpartner sector 
(or even without it~\cite{rnu1}) if SUSY breaking scale is lower than seesaw 
scale, a mass split between right and left neutrinos and sneutrinos will occur. 
As the right-hand neutrino super-field is assumed to be a singlet of the GUT 
gauge symmetry, it has only Yukawa-type of interaction. In such a setup $X$ 
can be a heavy right sneutrino decaying to a light sneutrino with the same 
leptonic number and a pair of Higgs or Higgsino~\cite{rnudm}. In place of 
assuming two $A$ particles in the final state we could consider them as being 
different $A$ and $A'$. But this adds a bit to the complexity of the model and 
does not change its general behaviour. For this reason we simply 
consider the same field. Diagram (\ref{decaymode}-b) is representative of a 
case where $X$ and $A$ are fermions, or $\Phi$ carries a conserved 
charge~\cite{snuaxion}. 

It was easier to consider a simple 3-vertex $X \rightarrow \Phi + A$ 
similar to what is considered in Ref.~\cite{warminf}. However, such a vertex 
does not always allow simultaneous conservation of energy. Moreover, as we 
will see later, more complex diagrams considered here will show how the 
decay mode affects the evolution of the condensate component.

The corresponding Lagrangians of these effective interactions are the 
followings:
\bea
{\mathcal L}_{\Phi} &=& \int d^4 x \sqrt{-g} \biggl [\frac{1}{2} g^{\mu\nu}
{\partial}_{\mu}\Phi {\partial}_{\mu}\Phi - \frac{1}{2}m_{\Phi}^2 {\Phi}^2 - 
\frac{\lambda}{n}{\Phi}^n \biggr ] \label{lagrangphi} \\
{\mathcal L}_{X} &=& \int d^4 x \sqrt{-g} \biggl [\frac{1}{2} g^{\mu\nu}
{\partial}_{\mu}X {\partial}_{\mu}X - \frac{1}{2}m_X^2 X^2 \biggr ] 
\label{lagrangx}\\
{\mathcal L}_{A} &=& \int d^4 x \sqrt{-g} \biggl [\frac{1}{2} g^{\mu\nu}
{\partial}_{\mu}A {\partial}_{\mu}A - \frac{1}{2}m_A^2 A^2 - 
\frac{\lambda'}{n'}A^{n'} \biggr ] \label{lagranga} \\
{\mathcal L}_{int} &=& \int d^4 x \sqrt{-g} \begin{cases} \mg \Phi X A^2, & 
\text{For (\ref{decaymode})-a} \\ \mg {\Phi}^2 XA, & \text{For 
(\ref{decaymode})-b} \end{cases}  \label{lagrangint}
\eea
In addition to the interaction between $X$, $\Phi$ and $A$ we have assumed a 
power-law self-interaction for $\Phi$ and $A$. If $A$ is a collective 
notation for other fields in the actual model, its self-interaction 
corresponds to the interaction between these unspecified fields. Again for 
the sake of simplicity in the rest of this work we consider $\lambda' = 0$. 
The unstable particle $X$ is assumed to have no self-interaction. 

Although the model presented here is quite general, for physical and 
observational reasons we concentrate on the case of a heavy $X$ particle 
as a candidate for the dark matter, $\Phi$ as a quintessence field, and 
interactions in (\ref{decaymode}) as candidate interactions for the decay 
of the dark matter and production of what is observed as dark energy. It is 
therefore necessary that $X$ and $\Phi$ have only a very weak interaction. 
Therefore couplings $\lambda$ and $\mg$ must be very small. 

In a realistic particle physics model, renormalization as well as 
non-perturbative effects can lead to complicated potentials for scalar fields. 
An example relevant for dark energy is a pseudo-Nambu-Goldston boson as $\Phi$ 
and potentials with a shift symmetry~\cite{quinpngb}. These models are 
interesting for the fact that the mass of the quintessence field does not 
receive quantum corrections and can be very small. Moreover, they can 
be easily implemented in SUSY theories in relation with right-neutrinos and 
sneutrinos (as candidate for $X$). The power-law potential considered here 
can be interpreted as the dominant term in the polynomial expansion of the 
potential. In any case, general aspects of the analysis presented here do not 
depend on the details of the particle physics and self-interaction, and can be 
applied to any model. The solution of the field equation and numerical 
quantities are however sensitive to the particle physics. The purpose of the 
present work is to investigate the behaviour of this model and to find what 
is most important for the formation and evolution of a condensate with 
characteristics similar to the observed dark energy. We leave the application 
of this analysis to realistic particle physics models to a future work and 
consider only simplest cases which are analytically tractable and permit 
exact or approximate analytical solutions. 

We decompose $\Phi (x)$ to a classical (condensate) and a quantum component:
\be
\Phi (x) = \varphi (x) + \phi (x) \quad \quad \langle \Phi \rangle \equiv 
\langle \Psi|\Phi|\Psi \rangle = \varphi (x) \quad \quad  \langle \phi \rangle 
\equiv \langle \Psi|\phi|\Psi \rangle = 0 \label{decomphi} 
\ee
Note that in (\ref{decomphi}) both classical and quantum components depend on 
the spacetime $x$. In studying inflation it is usually assumed that very 
fast expansion of the Universe washes out all the inhomogeneities and the 
condensed component is homogeneous. As we are studying the evolution after 
inflation, the distribution of unstable $X$ can have non-negligible 
inhomogeneities, specially if the decay is slow and perturbations have time 
to grow. 

We assume $\langle X \rangle = 0$ and $\langle A \rangle = 0$. Justification 
for these assumptions is the large mass and small coupling of $X$ which should 
reduce their number and their quantum correlation. In other words, when mass 
is large, the minimum of the effective potential for the classical component 
is pushed to zero (see (\ref{dyneffa}) and (\ref{dyneffb}) below). We find a 
quantitative justification for negligible condensation of massive fields in 
Sec. \ref{sec:classevol}. 

The Lagrangian of $\Phi$ is decomposed to:
\bea
{\mathcal L}_{\Phi} &=& {\mathcal L}_{\varphi} + {\mathcal L}_{\phi} + 
\int d^4 x \sqrt{-g} \biggl [\frac{1}{2}g^{\mu\nu}({\partial}_{\mu}\varphi 
{\partial}_{\nu}\phi + {\partial}_{\mu}\phi {\partial}_{\nu}\varphi) - 
m_{\Phi}^2\varphi\phi - \frac{\lambda}{n}\sum_{i=0}^
n \binom{n}{i}{\varphi}^i{\phi}^{n-i} \biggr ] + \nonumber \\
& & \int d^4 x \sqrt{-g} \begin{cases} \mg\varphi XA^2 +
\mg \phi XA^2 & \quad \quad \text{For (\ref{decaymode})-a} \\
\mg{\varphi}^2 XA + 2\mg\varphi \phi XA + \mg {\phi}^2 XA & \quad \quad 
\text{For (\ref{decaymode})-b} 
\end{cases} \label {lagrangeff}
\eea
Lagrangians ${\mathcal L}_{\varphi}$ and ${\mathcal L}_{\phi}$ are the same as 
(\ref{lagrangphi}) with respectively $\Phi \rightarrow \varphi$ and 
$\Phi \rightarrow \phi$. After replacing quantum terms by their expectation 
values, this Lagrangian leads to the following evolution equation for the 
condensate component:
\bea
\frac{1}{\sqrt{-g}}{\partial}_{\mu}(\sqrt{-g} g^{\mu\nu}{\partial}_{\nu}
\varphi) + m_{\Phi}^2 \varphi + \frac{\lambda}{n}\sum_{i=0}^{n-1} (i+1)
\binom{n}{i+1}{\varphi}^i\langle{\phi}^{n-i-1}\rangle - \mg \langle 
XA^2\rangle = 0 && \nonumber \\
\hspace {12cm} \text{For (\ref{decaymode})-a} && \label {dyneffa} \\ 
\frac{1}{\sqrt{-g}}{\partial}_{\mu}(\sqrt{-g} g^{\mu\nu}{\partial}_{\nu}
\varphi) + m_{\Phi}^2 \varphi + \frac{\lambda}{n}\sum_{i=0}^{n-1} 
(i+1)\binom{n}{i+1}{\varphi}^{i}\langle{\phi}^{n-i-1}\rangle - 2\mg\varphi 
\langle XA\rangle - 2\mg \langle \phi XA\rangle = 0 && \nonumber \\
\hspace {12cm} \text{For (\ref{decaymode})-b} && \label {dyneffb}
\eea
In the case of the interaction mode (\ref{decaymode}-a), the interaction 
Lagrangian depends linearly on $\Phi$ and appears as an external source 
in the field equation (\ref{dyneffa}). For both decay modes, 
if $n \geqslant 2$, the term $i = 0$ in the sum of the self-interaction 
terms also contributes to the non-homogeneous component, and the term 
$i = 1$ contributes to the effective mass of the classical field $\varphi$. 
Note that what we call non-homogeneous component or 
{\it external source terms} have in fact implicit dependence on 
$\varphi$ . The reason is the coupling between quantum interactions and the 
evolution of the classical component. We will show later that 
these terms play the role of a feedback between production and evolution of 
the condensate. In fact, (\ref{decaymode}-b) has a reach structure and various 
evolution histories are possible depending on the value and sign of $\mg$, 
the coupling to $X$, self-coupling $\lambda$, and the order of the 
self-interaction potential $n$. For instance, the mass can become imaginary 
(tachyonic) even without self-interaction leading to symmetry breaking. 
Tachyonic scalar fields have been suggested as quintessence field specially 
in the framework of models with $w < -1$~\cite{tachyonquin}. The decay mode 
(\ref{decaymode}-a) by contrast has a field equation very similar 
to the classical model studied in Ref.~\cite{houridmquin}. We discus in detail 
the differences of these decay modes in the next sections. 

In addition to interaction terms in the Lagrangian of the total field $\Phi$, 
the Lagrangian of the purely quantic component $\phi$ includes terms depending 
on the classical component $\varphi$. Derivative term 
$\frac{1}{2}g^{\mu\nu}({\partial}_{\mu}\varphi 
{\partial}_{\nu}\phi + {\partial}_{\mu}\phi {\partial}_{\nu}\varphi)$, mass 
term $m^2\varphi\phi$, $i = n-1$ term in the self-interaction, and 
$2\mg \varphi\phi XA$ are linear in $\phi$ and only affect the 
renormalization of the propagator~\cite{infrenorm}. For $n \geqslant 2$ the 
term $i = n - 2$ in the self-interaction sum contributes in the effective 
mass of $\phi$ and makes it time dependent.

We use Schwinger closed time path (also called in-in) formalism to calculate 
expectation values. Recent reviews of this formalism are 
available~\cite{ctprev} 
and here we only present the results. Zero-order (tree) diagrams for the 
expectation values (\ref{dyneffa}) and (\ref{dyneffb}) are shown in 
(\ref{xaadiag}), (\ref{xadiag}) and (\ref{xaphidiag}). The next 
relevant diagrams are of order $g^3$ and negligible for the dark energy model. 
One example of higher order diagrams is shown in (\ref{xaadiag}). These types 
of diagrams are specially important for studying renormalization in the 
context of a realistic particle physics model. Thus for the phenomenological 
models considered here, we can ignore them.
\bea
g\varphi \langle XA^2\rangle = \quad  
\parbox{40mm}{\begin{fmfgraph*}(30,20)
\fmfleft{i1}
\fmfright{o2,o3,o4}
\fmf{dashes,label=$\varphi$}{i1,v1}
\fmf{plain,label=$A$}{v1,o2}
\fmf{dbl_plain}{v1,v2}
\fmf{dbl_plain,label=$X$}{v2,o3}
\fmf{plain,label=$A$}{v1,o4}
\fmfdot{v1}
\end{fmfgraph*}} + && 
\parbox{40mm}{\begin{fmfgraph*}(30,30)
\fmfleft{i1}
\fmfright{o2,o3,o4}
\fmf{dashes,label=$\varphi$}{i1,v1}
\fmf{dbl_plain,label=$X$}{v3,o2}
\fmf{dbl_plain,label=$X$,label.side=left}{v1,v2}
\fmf{plain,label=$\phi$,label.side=left}{v2,v3}
\fmf{plain}{v2,v4}
\fmf{plain,label=$A$,label.side=left}{v4,o3}
\fmf{plain,label=$A$,label.side=left}{v2,o4}
\fmf{plain,label=$A$,left=0.5,tension=1/3}{v3,v1}
\fmf{plain,right=0.5,tension=1/3}{v3,v1}
\fmfdot{v1}
\fmfdot{v2}
\fmfdot{v3}
\end{fmfgraph*}} + \ldots \label {xaadiag}
\eea
\medskip
\bea
g{\varphi}^2 \langle XA\rangle = \quad  
\parbox{40mm}{\begin{fmfgraph*}(30,20)
\fmfleft{i1,i2}
\fmfright{o3,o4}
\fmf{dashes,label=$\varphi$}{i1,v1}
\fmf{dashes,label=$\varphi$}{i2,v1}
\fmf{plain,label=$A$}{v1,o3}
\fmf{dbl_plain,label=$X$}{v1,o4}
\fmfdot{v1}
\end{fmfgraph*}} + \ldots && \label {xadiag} \\
g\varphi \langle \phi XA\rangle = \quad  
\parbox{40mm}{
\begin{fmfgraph*}(30,20)
\fmfleft{i1}
\fmfright{o2,o3,o4}
\fmf{dashes,label=$\varphi$}{i1,v1}
\fmf{dbl_plain,label=$X$}{v1,o2}
\fmf{plain}{v1,v2}
\fmf{plain,label=$\phi$}{v2,o3}
\fmf{plain,label=$A$}{v1,o4}
\fmfdot{v1}
\end{fmfgraph*}} + \ldots \label {xaphidiag}
\eea
Diagrams for self-interaction terms in (\ref{dyneffa}) and (\ref{dyneffb}) 
are similar to (\ref{xadiag}) with $n-1$ external lines of type 
$\phi$ or $\varphi$. The dash line presents the classical component of 
$\Phi$. The corresponding expectation values at zero order are:
\bea
g\varphi \langle XA^2\rangle & = & g \varphi (x) \int \sqrt{-g} d^4y 
\biggl [G_A^> (x,y) G_A^> (x,y) G_X^> (x,y) - G_A^< (x,y) 
G_A^< (x,y) G_X^< (x,y)\biggr ] \label {valxaa}\\
g{\varphi}^2 \langle XA\rangle & = & g{\varphi}^2 \int \sqrt{-g} d^4y 
\biggl [G_A^> (x,y) G_X^> (x,y) - G_A^< (x,y) G_X^< (x,y)\biggr ] 
\label {valxa} \\
g\varphi \langle \phi XA\rangle & = & g \varphi (x) \int \sqrt{-g} d^4y 
\biggl [G_{\phi}^> (x,y) G_A^> (x,y) G_X^> (x,y) - 
G_{\phi}^< (x,y) G_A^< (x,y) G_X^< (x,y)\biggr ] \label {valxaphi}
\eea
Future and past propagators $G^>$ and $G^<$ are defined as:
\bea
G^> (x,y) &\equiv& -i \langle \psi (x) {\psi}^{\dagger}(y) \rangle = 
-i \tr (\psi (x) {\psi}^{\dagger}(y)\rho) \label{proggt}\\
G^< (x,y) &\equiv& \mp i \langle {\psi}^{\dagger} (y) \psi (x) \rangle = 
\mp i \tr ({\psi}^{\dagger} (y) \psi (x)\rho)
\label{progls}
\eea
where $\psi (x)$ presents one of $\phi$, $X$ or $A$ fields and 
$\rho = |\Psi\rangle\langle\Psi|$ is the density (projection) operator for the 
state $|\Psi\rangle$. The upper and lower signs in (\ref{progls}) are 
respectively for bosons and fermions. Definitions (\ref{proggt}) and 
(\ref{progls}) correspond to the general case of a complex field. Here we only 
consider real fields and therefore $\psi (x) = {\psi}^{\dagger}(x)$. Feynman 
propagators are related to $G^> (x,y)$ and $G^< (x,y)$:
\bea
G_F (x,y) &\equiv& -i \langle T\psi (x) {\psi}^{\dagger}(y) \rangle = 
G^> (x,y) \Theta (x^0-y^0) + G^< (x,y) \Theta (y^0-x^0)\label{progf}\\
\bar{G}_F (x,y) &\equiv& -i \langle \bar{T}\psi (x) {\psi}^{\dagger}(y)
\rangle = G^> (x,y) \Theta (y^0-x^0) + G^< (x,y) \Theta (x^0-y^0)
\label{progfbar}
\eea
The next step is the calculation of propagators.

\subsection {Propagators in an expanding universe} \label{sec:prog}
Feynman propagators $G^i_F (x,y), ~i = \phi,~X,~A~$ can be determined using 
field equations from Lagrangians (\ref{lagrangphi})-(\ref{lagrangint}). As the 
concept of Green's functions is only applicable to the linear differential 
equations, we have to linearize the field equations and treat interactions 
perturbatively, i.e. as quantum corrections. Free propagators of $\phi$, $X$ 
and $A$ are:
\bea
\frac{1}{\sqrt{-g}}{\partial}_{\mu}(\sqrt{-g} g^{\mu\nu}{\partial}_{\nu}G^{\phi}_F 
(x-y)) + (m_{\Phi}^2 + (n-1) \lambda {\varphi}^{n-2}) G^{\phi}_F (x-y) & = & 
-i \frac{{\delta}^4 (x-y)}{\sqrt{-g}} \label {propagphi} \\
\frac{1}{\sqrt{-g}}{\partial}_{\mu}(\sqrt{-g} g^{\mu\nu}{\partial}_{\nu}G^i_F 
(x-y)) + m_i^2 G^i_F (x-y) & = & -i\frac{{\delta}^4 (x-y)}{\sqrt{-g}} \quad 
i = X, A \label {propagx}
\eea
The free propagator of $\phi$ is independent of the type of interaction with 
$X$ and $A$ and therefore, equation (\ref{propagphi}) is valid for both 
interaction models presented in (\ref{decaymode}). Note also that 
$G^{\phi}_F (x-y)$ is coupled to the classical field $\varphi$ even at the 
lowest quantum perturbation order. On the other hand, evolution equations 
(\ref{dyneffa}) and (\ref{dyneffb}), depend on the interaction 
between quantum fields $\phi$, $X$ and $A$. Therefore, this model is coupled 
at all orders.

To proceed, we neglect the effect of spatial anisotropy and consider a flat 
homogeneous metric with synchronous or conformal time:
\be
ds^2 = dt^2 - a^2(t){\delta}_{ij}dx^idx^j = a^2(\eta) (d{\eta}^2 - 
{\delta}_{ij}dx^idx^j) \quad\quad dt \equiv a d{\eta} \label{metric}
\ee
where $t$ and $\eta$ are respectively comoving and conformal time. It is 
more convenient to write evolution and propagator equations with respect to 
conformal time $\eta$. After a variable change:
\be
\chi \equiv a\varphi \quad \Upsilon \equiv a\phi \quad \cx \equiv aX \quad 
\ca \equiv aA
\ee
and by using the metric (\ref{metric}), the evolution equation of the 
classical field $\chi$ takes the following form:
\bea
{\chi}'' - \delta_{ij}\partial_i\partial_j \chi + (a^2 m_{\Phi}^2 - 
\frac{a''}{a}) \chi + \frac{\lambda a^{n-4}}{n}\sum_{i=0}^{n-1} (i+1) 
\binom{n}{i+1} {\chi}^i \langle {\Upsilon}^{n-i-1}\rangle - \mg \langle 
\cx\ca^2 \rangle = 0 && \nonumber\\
\hspace{12cm} \text{For (\ref{decaymode})-a} && \label{evola} \\ 
{\chi}'' - \delta_{ij}\partial_i\partial_j \chi + (a^2 m_{\Phi}^2 - 
\frac{a''}{a}) \chi + \frac{\lambda a^{n-4}}{n}\sum_{i=0}^{n-1} (i+1) 
\binom{n}{i+1} {\chi}^i\langle {\Upsilon}^{n-i-1}\rangle - 2\mg \chi \langle 
\cx\ca \rangle - 2\mg \langle \Upsilon \cx\ca \rangle = 0 && \nonumber\\
\hspace{12cm} \text{For (\ref{decaymode})-b} && \label {evolb}
\eea
Propagator of quantum fields $\Upsilon$, $\cx$ and $\ca$ are:
\bea
&&\frac{d^2}{d{\eta}^2} G^{\Upsilon}_F (x,y) - \delta_{ij}\partial_i\partial_j 
G^{\Upsilon}_F (x,y) + (a^2 m_{\Phi}^2 - \frac{a''}{a} + (n-1) \lambda 
a^2{\varphi}^{n-2}) G^{\Upsilon}_F (x,y) =  
-i \frac {{\delta}^4 (x-y)}{a} \label {propup}\\
&&\frac{d^2}{d{\eta}^2} G^{\mathtt i}_F (x,y) - \delta_{ij}\partial_i\partial_j 
G^{\mathtt i}_F (x,y) + (a^2 m_{\mathtt i}^2 - \frac{a''}{a}) G^{\mathtt i}_F 
(x,y) = -i \frac {{\delta}^4 (x-y)}{a} \quad {\mathtt i} = \cx, \ca 
\label {propcx}\\
&& G^{\Upsilon}_F = a(\eta)G^{\phi}_F \quad , \quad G^{\cx}_F = a(\eta) G^X_F 
\quad , \quad G^{\ca}_F = a(\eta) G^A_F \label {progredef}
\eea
with $f' \equiv df/d\eta$. Note that the classical component of $\Phi$ appears 
as a spacetime dependent mass term for its quantum component $\phi$ (or 
equivalently $\Upsilon$). Fourier transform can be applied to (\ref{propcx}), 
but the spacetime dependence of coefficients in (\ref{evola}), (\ref{evolb}) 
and (\ref{propup}) makes this method useless for solving these equations. 
Nonetheless, if these terms are small and/or vary slowly, one can first ignore
them and solve the equations. Then, by applying the WKB method, a more precise 
solution can be obtained. Equations (\ref{propup})-(\ref{propcx}) are second 
order partial differential equations, and therefore propagators are linear 
combination of two independent solutions of the associated homogeneous 
equations. The delta function on the right hand side however leads to a 
discontinuity which appears as a consistency condition for the solutions and 
fixes the ambiguities in the propagator solution. This will be described in 
details in the rest of this section.

$X$ particles are presumably produced during reheating epoch~\cite{sdmprod} 
and begin their decay afterward. In this epoch relativistic particles dominate 
the density of the Universe. Thus, we first consider this epoch. Fortunately, 
for this epoch the homogeneous field equation has an exact solution. In matter 
dominated epoch only for special cases an analytical solution exists. 
They are discussed in the Appendix \ref{app:c}. The expansion in the radiation 
domination epoch has the following time dependence:
\be
a = a_0 \biggl (\frac{t}{t_0} \biggr )^{\frac{1}{2}} = a_0 \frac{\eta}{\eta_0} 
\label {expan}
\ee
And thus $a'' = 0$. After taking the Fourier transform of the spatial 
coordinates and neglecting the $\varphi$-dependent term, the solutions of 
the associated homogeneous equation of (\ref{propup}) and (\ref{propcx}) are 
well known~\cite{qmcurve}~\cite{integbook}:
\bea
&& (\frac{d^2}{d{\eta}^2} + k^2 + a^2 m_i^2) \um_k^i (\eta) = 0 \label{eqhomo} \\
&& \um_k^i = \int d^3 \vec{x} \um^i (x) e^{i\vec{k}.\vec{x}} \quad i = 
\Phi~,~\cx~,~\ca \label {fourier}\\
&& \um_k^i (\eta) = c_k^i D_{q_i}(\alpha_i \eta) + d_k^i D_{q_i}(-\alpha_i \eta) 
\label{homosol} \\
&& \alpha_i \equiv (1+i)\sqrt{B_i} \quad q_i \equiv - 
\frac{1+\frac{ik^2}{B_i}}{2} \quad B_i \equiv \frac{a_0 m_i}{\eta_0} = 
\frac {m_i}{H_0 \eta_0^2} = a^2_0 H_0 m_i \label{homosolparam}
\eea
where $a_0$ and $H_0$ are respectively expansion factor and Hubble constant at 
the initial conformal time $\eta_0$. The function $D_q(z)$ is the parabolic 
cylindrical function. From now on for simplicity we drop the species index $i$ 
unless when its presence is necessary. We call two independent solutions 
of (\ref{eqhomo}) in a general basis $\um_k$ and ${\mathcal V}_k$. If we want 
these solutions correspond to the coefficients of the canonical decomposition 
of $\phi$ ((\ref{canon}) in Appendix \ref{app:b}), we must choose a basis such 
that ${\mathcal V}_k = \um_k^*$. In the rest of this work we only consider 
this basis. The corresponding equation for the free Feynman propagator 
(free 2-point Green's function) is:
\be
(\frac{d^2}{d{\eta}^2} + k^2 + a^2 m_i^2) G_k (\eta,\eta') = 
-i \frac {\delta (\eta - \eta')}{a} \label{prophomo}
\ee
When $\eta \neq \eta'$, (\ref{prophomo}) is the same as (\ref{eqhomo}) and 
therefore solutions of the former is a linear combination of two independent 
solutions of the latter. According to the definition of Feynman propagators 
(\ref{proggt}) and (\ref{progls}) it can be divided to past and future 
propagating components $G^<$ and $G^>$. With $\eta \leftrightarrow \eta'$ 
these propagators change their role: $G^< \leftrightarrow G^>$. Therefore, 
$G (\eta, \eta')$ has the following expansion:
\be
iG (\eta, \eta') = \biggl [{\mathcal A}^>_k \um_k (\eta)\um^*_k (\eta') + 
[{\mathcal B}^>_k \um^*_k (\eta) \um_k (\eta')\biggr ] \Theta (\eta - \eta') + 
\biggl [{\mathcal A}^<_k \um_k (\eta)\um^*_k (\eta') + 
[{\mathcal B}^<_k \um^*_k (\eta) \um_k (\eta')\biggr ] \Theta (\eta' - \eta) 
\label{progexpand}
\ee
where ${\mathcal A}^>_k$, ${\mathcal B}^>_k$, ${\mathcal A}^<_k$ and 
${\mathcal B}^<_k$ are integration constants. In the Appendix \ref{app:b} 
we show that for the free propagators i.e. at the lowest perturbation order, 
if the state $|\Psi\rangle$ is not vacuum, it is possible to 
include its effect in the boundary conditions imposed on the propagator. 
Comparing (\ref{progexpand}) with (\ref{propst}) in the Appendix \ref{app:b}, 
the relation between these constants and the initial state can be concluded:
\bea
&&{\mathcal A}^>_k = 1 + {\mathcal B}^>_k \quad , \quad {\mathcal B}^<_k = 
1 + {\mathcal A}^<_k \label{abrel} \\
&&{\mathcal A}^<_k = {\mathcal B}^>_k = \sum_i \sum_{k_1 k_2 
\ldots k_n} \delta_{kk_i} |\Psi_{k_1 k_2 \ldots k_n}|^2 \label{abpsi}
\eea
It is easy to see that with these relations the consistency condition defined 
as:
\be
G^> (\eta,\eta')\biggl |_{\eta = \eta'} = G^< (\eta,\eta')\biggl |_{\eta = 
\eta'} \label {consistcond} 
\ee
is automatically satisfied. Therefore propagators over a non-vacuum state 
$\Psi$ only depends on this state and the solutions of the 
field equation considered as {\it the free particle states}. On the other 
hand, these solutions depend on two arbitrary constants $c_k^i$ and $d_k^i$ 
which should be fixed by initial conditions. We have used decomposition 
(\ref{progexpand}) along with canonical decomposition of the quantum field 
(see (\ref{canon}) to (\ref{psistate}) in Appendix B)to fix the integration 
constants $c_k^i$ and $d_k^i$. The advantage of this method is that it 
explicitly relate dynamical constants to the physical properties of 
environment in which the quantum field is living.

There is one more consistency condition that propagators should satisfy. 
By integrating two sides of equation (\ref{prophomo}) with respect to $\eta$ 
in an infinitesimally region around $\eta'$ we find the following constraint:
\be
\um_k^{'} (\eta) \um^*_k (\eta) - \um_k (\eta) \um^{'*}_k (\eta)= 
\frac {-i}{a(\eta)} \label{derivcond}
\ee
This relation fixes one of the dynamical constants in (\ref{homosol}). It can 
be chosen to be the normalization of the propagator. It is however 
interesting to note that multiplying both sides of (\ref{derivcond}) with an 
arbitrary constant rescale $a(\eta)$ which is equivalent to redefinition of 
$a_0$. Rescaling of $a_0$ is equivalent to redefinition of coordinates and 
therefore is not an observable. In Minkovsky spacetime the scale factor $a$ 
is fixed to 1, and therefore there is no place for rescaling. Thus in a 
Minkovsky space the normalization of the propagators is an observable and 
affecting final results. The scaling properties in FLRW or De-Sitter metric is 
a consequence of diffeomorphism invariance in the framework of curved 
spacetimes and general relativity.

\subsection{Initial Conditions} \label{sec:initcond}
Field equations are second order differential equations and need the initial 
value of the field and its derivative or a combination of them to be 
totally described. The general initial conditions for a bounded system, 
including both Neumann and Dirichlet conditions as special cases, are the 
followings~\cite{qftinit}:
\be
n^\mu \partial_\mu \um = i{\mathcal K} \um \quad ,\quad 
n^\mu \partial_\mu \um^* = -i{\mathcal K}^* \um^* \quad ,\quad g_{\mu\nu}n^\mu 
n^\nu = 1 
\label{geninit} 
\ee
The 4-vector $n^\mu$ is the normal to the boundary surface. If the boundary is 
space-like, the normal $n^\mu$ can be normalized to $a^{-1}(1,0,0,0)$, then:
\bea
a^{-1}\partial_{\eta} \um = i {\mathcal K} \um \quad , \quad a^{-1}
\partial_{\eta} \um^* = -i {\mathcal K}^* \um^* \label{geninitorth} 
\eea
 Constants ${\mathcal K}_i$ and ${\mathcal K}_f$ depends on $k$. In a general 
boundary problem the boundary conditions must be defined for all 
the boundaries. Thus, in a cosmological setup the initial conditions 
(\ref{geninitorth}) must be applied to a past (initial) and future (final) 
boundary surfaces. This is the strategy suggested in Ref.~\cite{qftinit}. 
The past 
and future boundary conditions are respectively applicable only to past and 
future propagators. Assuming different values for ${\mathcal K}$ on the past 
and future boundary, one finds:
\be
{\mathcal K}_j = -i\frac {\um^{'}_k (\eta_j)}{a_j \um_k(\eta_j)} \quad , 
\quad j = i, f 
\label {alphavac}
\ee
Indexes $i$ and $f$ refer to the value of quantities at initial and final 
boundary conditions. These boundary conditions relate ${\mathcal K}_i$ and 
${\mathcal K}_f$ 
to $c_k$ and $d_k$ in (\ref{homosol}). In a cosmological context although 
${\mathcal K}_f$ apriori can be decided based on observations, the value 
of ${\mathcal K}_i$ is unknown and leaves one arbitrary constant in the 
solution or it should be selected according to a special model considered for 
the physics of early universe. This arbitrariness of the general solution, 
or in other words the vacuum of the theory, is well known~\cite{vacuu}. In the 
case of inflation, this leads to a class of possible vacuum solutions called 
$\alpha$-vacuum. For instance if:
\be 
{\mathcal K}_i = {\mathcal K}_f = \sqrt {k^2/a^2_i + m^2} \label{bunchdavies}
\ee
one obtains the well known Bunch-Davies solution~\cite{qftinit}. 

Another way of proceeding is using the consistency condition (\ref{derivcond}) 
to fix one of the arbitrary constants and applying the boundary condition 
(\ref{alphavac}) only to one of the initial or final 3-surfaces. Although this 
does not solve the problem of arbitrariness of ${\mathcal K}$ and its $k$ 
dependence, it reduces it to only one of the boundary surfaces, for instance 
to the final 3-surface, and make the choice of (\ref{bunchdavies}) physically 
motivated. Besides, in fixing only one of the boundary conditions, the 
causality of the solution is transparent - the state of the second boundary 
is directly related on the choice of the first one through the evolution 
equation. 

Another and somehow hidden arbitrariness in this formalism is the fact that 
apriori $k$ dependence of the boundary constant ${\mathcal K}$ does not need 
to be the same for all the fields of the model. However, different $k$ 
dependence for the boundary conditions breaks the Equivalence Principal. 
Similarly, a value different from (\ref{bunchdavies}) for ${\mathcal K}$ will 
lead to the breaking of the translation 
symmetry~\cite{qftinit}~\cite{infrenorm}. In the context of quantum gravity 
the violence of both of these laws are expected and therefore, in a general 
framework they should be considered.

\subsection{WKB approximation and back reactions} \label{sec:wkbprog}
Finally after finding the solution of linearized field equation (\ref{eqhomo}) 
and corresponding propagator (\ref{prophomo}) for the field $\Upsilon$, we 
should add the effect of spacetime dependence of the mass term. According to 
the WKB prescription we must replace $\alpha_\Phi \eta$ in $\um$ with:
\be
\alpha_\Phi \eta \rightarrow \alpha_\Phi \int d\eta \biggl (1 + (n-1) 
\lambda \frac{\varphi_k^{n-2}(\eta)}{m_{\Phi}^2} \biggr )^{\frac{1}{4}} 
\label{wkbappr}
\ee
where $\varphi_k$ is the Fourier transform of $\varphi (x)$ and $\alpha_\Phi$ 
is defined in (\ref{homosolparam}). With this correction apparently 
we have the solution for all the propagators at lowest order. However, 
$\varphi (x)$ (or equivalently $\chi$) evolves according to equation 
(\ref{evola}) or (\ref{evolb}) that depend on the expectation values of the 
quantum fields, specially $\phi$. Therefore the propagators and the condensate 
are coupled even at lowest order, and only through a numerical calculation 
a final solution for each of them can be obtained. In fact the reason for 
coupling of classical and quantum fields is the self-coupling of $\Phi$. If 
the self-interaction of $\Phi$ is negligible, the solutions of evolution 
equation of the classical component (\ref{evola}) or (\ref{evolb}) 
are similar to the field equation of the quantum component. In presence of a 
self-interaction however this equation is non-linear and must be solved 
numerically.

In addition to the self-coupling, the state $|\Psi\rangle$ for which we 
calculate propagators and expectation values is a source of back-reaction 
(classical effects of back-reaction are studied in Ref.~\cite{backreact}). 
It defines the quantum state of the Universe at the time when the 
interactions or more exactly the decay of $X$ is studied. However, 
$|\Psi\rangle$ evolves due to interactions between species and the expansion 
of the Universe. They are responsible for the variation of the number of 
particles and their momentum distribution and therefore evolution of 
$|\Psi\rangle$. In fact, in the path-integral formulation of closed 
path integral, a non-vacuum state $|\Psi\rangle$ adds a functional integral to 
the path-integral~\cite{ctprev} that presents the projection of the state on a 
predefined basis. When the Fock space is evolving, as it is the case in the 
cosmological context, the projection of $|\Psi\rangle$ evolves and path 
integrals become unfactorizable. 

As an example we consider an initial non-zero distribution for the $X$ 
particles and no $A$ or $\Phi$ particles. At a later time, the number of 
$X$ particles in $|\Psi\rangle$ is reduced due to their decay and the 
expansion of the Universe. Their temperature or mean kinetic energy if they 
don't have a thermal distribution also decreases. On the other hand, 
a non-zero number of $\Phi$ and $A$ particles are created. The latter at 
their production are relativistic and non-thermal. But if they have 
self-interaction and/or interaction with other fields which we ignored here, 
their momentum distribution will change both by interactions and due to the 
expansion of the 
Universe. This means that $|\Psi\rangle$ will change and its variation is 
reflected on the evolution of the classical component $\varphi$, the quantum 
component $\phi$, the expansion factor $a (\eta)$, and the thermalization of 
$A$ which we assume to have interactions with other particles (see also 
Appendix \ref{app:b}).

\section{Evolution of the classical field without self-interaction} 
\label{sec:classevol}
To get an insight into the evolution of the classical component $\varphi$, 
in this section we neglect self-interaction of $\Phi$ which is the main source 
of the non-linearity and coupling of the equations. We find an analytical 
solution for the evolution of condensate and discuss the difference between 
two decay modes, as well as the effect of the other parameters. The evolution 
of $|\Psi\rangle$ is introduced by a simple parametrization. We discuss its 
effect and determine the range in which the condensate can have a behaviour 
similar to the dark energy.

\subsection{Expectation values} \label{sec:expecval}
Neglecting the self-interaction of $\Phi$, expression (\ref{homosol}) is 
the exact solution of the field equation and (\ref{progexpand}) is the exact 
propagator. Therefore, we can calculate the expectation values (\ref{valxaa}) to 
(\ref{valxaphi}):
\bea
g \langle \cx \ca^2\rangle (x) & = & \frac{-ig}{(2\pi)^6} 
\int d^3 k_1 d^3 k_2 d^3 k_3 e^{-i\vec{x}.(\vec{k}_1 + \vec{k}_2 + \vec{k}_3)} 
\delta^{(3)} (\vec{k}_1 + \vec{k}_2 + \vec{k}_3) \int d\eta'\sqrt{-g} 
\nonumber \\ 
&&\biggl [G_{k_1}^{\ca >} (\eta,\eta') G_{k_2}^{\ca >} (\eta,\eta') 
G_{k_3}^{\cx >} (\eta,\eta') - G_{k_1}^{\ca <} (\eta,\eta') G_{k_2}^{\ca <} 
(\eta,\eta') G_{k_3}^{\cx <} (\eta,\eta') \biggr ] \label {valxaacal}\\
g \langle \cx \ca\rangle (x) & = & \frac{-ig}{(2\pi)^3}
\int d^3 k_1 d^3 k_2 e^{-i\vec{x}.(\vec{k}_1 + \vec{k}_2)} \delta^{(3)} 
(\vec{k}_1 + \vec{k}_2) \int d\eta'\sqrt{-g} \nonumber \\
&& \biggl [G_{k_1}^{\ca >} (\eta,\eta') G_{k_2}^{\cx >} (\eta,\eta') - 
G_{k_1}^{\ca <} (\eta,\eta') G_{k_2}^{\cx <} (\eta,\eta') \biggr ]
\label {valxacal} \\
g \langle \Upsilon \cx \ca \rangle (x) & = & \frac{-ig}{(2\pi)^6} 
\int d^3 k_1 d^3 k_2 d^3 k_3 e^{-i\vec{x}.(\vec{k}_1 + \vec{k}_2 + \vec{k}_3)} 
\delta^{(3)} (\vec{k}_1 + \vec{k}_2 + \vec{k}_3) \int d\eta'\sqrt{-g} 
\nonumber \\
&& \biggl [G_{k_1}^{\Upsilon >} (\eta,\eta') G_{k_2}^{\ca >} (\eta,\eta') 
G_{k_3}^{\cx >} (\eta,\eta') - G_{k_1}^{\Upsilon <} (\eta,\eta') 
G_{k_2}^{\ca <} (\eta,\eta') G_{k_3}^{\cx <} (\eta,\eta') \biggr ] 
\label {valxaphical}
\eea
The most important aspect of these integrals for us is their time dependence 
because these expectation values contribute to the build-up and the time 
evolution of the condensate. Although the spatial spectrum is important 
specially for observational verification of the model, its role is secondary 
and comes later. 

Regarding (\ref{homosol}) and (\ref{progexpand}), it is easy to conclude that 
time integrals in (\ref{valxaacal})-(\ref{valxaphical}) are similar to the 
following:
\be
{\mathcal I}(\eta) \equiv \prod_{i=1}^N D_{q^*_i} (\gamma_i z) 
\int_{\eta_0}^{\eta} d\eta 
\sqrt{-g} \prod_{i=1}^N D_{q_i} (\beta_i z') \quad \quad q_i \equiv - 
\frac{1\pm\frac{ik^2}{B_i}}{2} \quad z' \equiv \alpha_i \eta' \quad \beta_i \in 
\{\pm,~\pm i\} \quad \gamma_i = -i \beta^*_i \label{prointeg}
\ee
where $\alpha_i$ and $B_i$ are defined in (\ref{homosolparam}) and $N$ is the 
number of fields in the expectation value brackets. For simplicity we use the 
product index in (\ref{prointeg}) for distinguishing the field species too. 
The constant coefficients of these integrals are of the form $\prod_i A^i_k 
{\mathcal C}^i_k {\mathcal C}^{i*}_k$ with $A^i_k \in \{{\mathcal A}^{i>}_k,
{\mathcal B}^{i>}_k,{\mathcal A}^{i<}_k,{\mathcal B}^{i<}_k\}$, 
${\mathsf C}^i_k \in \{c^i_k, d^i_k\}$ and ${\mathsf C}^{i*}_k \in \{c^{i*}_k, 
d^{i*}_k\}$ for the corresponding species. These integrals don't have 
analytical solutions. Using the asymptotic properties of the parabolic 
cylindrical functions, we estimate the late time behaviour of 
these integrals\footnote{Validity of both approximations depends on the 
spatial scale $k_i$ and the constant $B_i$ for each species. The initial 
time $\eta_0$ and Hubble constant $H_0$ also play important roles in these 
approximations and in the general behaviour of the model. Here we take these 
conditions for granted.}:
\bea
&& D_q (z) \sim e^{-\frac{z^2}{4}} z^q \biggl [1 - \frac{q(q-1)}{2z^2} + 
\ldots \biggr ]\quad , \quad |z| \gg 1 \quad , \quad |z| \gg |q| 
\label {dapproxl} \\
&& D_q (z) \sim 2^q e^{-\frac{z^2}{2}} \biggl [\frac{\sqrt{\pi}}{\Gamma 
(\frac {1-q}{2})} - \frac{\sqrt{2\pi} z}{\Gamma (\frac {-q}{2})} + \ldots 
\biggr] \quad , \quad |z| \ll 1 \label {dapproxs}
\eea
The validity of these regimes depends on the mass of the corresponding field 
and on the cosmological parameters at the initial time $\eta_0$:
\be
|\beta_i z_0| = \sqrt{\frac{2m_i}{H_0}} \label{initz}
\ee
For a light $\Phi$, its mass at the initial time $\eta_0$ can be comparable or 
lighter than the Hubble constant $H_0$ and therefore approximation 
(\ref{dapproxs}) is applicable. As for $X$ and $A$, it seems unlikely that in 
any relevant particle physics model for these fields $\frac{2m_i}{H_0}$ be 
small and therefore approximation (\ref{dapproxl}) must be applied. 

Applying these approximations to (\ref{prointeg}), at lowest order in $\eta$ 
this integral has the following time dependence:
\bea
&& {\mathcal I}(\eta) \sim \biggl ( \frac{a_0}{\eta_0} \biggr )^4 F (k)
\eta^{5 - N'}e^{\frac {i}{2}\sum_i \beta^2_i B_i \eta^2} \label{appint} \\
&& F (k) \propto \frac {\prod_j \frac {sh \biggl (\pi \frac{k^2_j}
{B_j}\biggr )}{\frac{k^2_j}{B_j}}}{5 - N' + \frac{i}{2}\sum_{i'} 
\frac{k^2_j}{B_{i'}}} \label{appk}
\eea
where $N'$ is the number of fields to which approximation (\ref{dapproxl}) can 
be applied. Index $i$ refers to all the fields. Indexes $i'$ refers to fields 
for which approximation (\ref{dapproxl}) is applicable and $j$ for ones with 
(\ref{dapproxs}) approximation. If $j =0$, the nominator in (\ref{appk}) is 1. 
In this case the effect of the fields for which $|\beta z|$ in the 
integration interval is small contributes only in the oscillating term of 
(\ref{appint}) and in the higher-order terms of the polynomial expansion.

For estimating the expectation values (\ref{valxaacal})-(\ref{valxaphical})
in addition to time dependence included in integral (\ref{prointeg}) one has 
to take into account the time variation of the state $|\Psi\rangle$ both due 
to the expansion of the Universe and due to the back-reaction of the decay on 
the density of particles. Neglecting the effect of the decay on the density 
for a short duration after the massive production of $X$ particles during 
reheating, the main source of the time evolution of $|\Psi\rangle$ and 
$|\Psi_{k_1 k_2 \ldots k_n}|^2$ is the expansion. Assuming a very heavy $X$, 
it can be considered as non-relativistic at the time of its production and 
its density decreases by $a^3 \propto \eta^3$. The other two fields $A$ and 
specially $\Phi$ are relativistic and their density decreases by a factor of 
$a^4 \propto \eta^4$. 

As the lifetime of $X$ particles is very long and thus $g$ is very small, 
the contribution of $A$ and $\phi$ particles to the total number, energy 
density, and entropy of the relativistic matter at this epoch is very small. 
Therefore, one expects that after an initial fast increasing trend of the 
expectation values in (\ref{valxaacal})-(\ref{valxaphical}), they will slow 
down and probably approach a constant or begin to decrease. The turning point 
depends on the expansion rate, lifetime of $X$ particles or equivalently the 
coupling constant $g$, the density of $X$ particles at the end of reheating 
through its effect on $|\Psi_{k_1 k_2 \ldots k_n}|^2$, and finally the mass 
of $\Phi$ and $A$ particles that can be estimated from (\ref{appint}). 
Larger the number of fields with $m/H_0 \gg 1$, slower 
the time evolution of (\ref{appint}) and sooner the turning point of the 
expectation values. This observation is consistent with our initial assumption 
of no condensation for $X$ and $A$. We expect that only for the field $\Phi$ 
there is a time interval in which $m_{\Phi}/H_0 < 1$. Thus, at late times 
$\langle \cx \ca^2\rangle (x)$ decreases faster than 
$\langle \cx \ca\rangle (x)$ and $\langle \Upsilon \cx \ca \rangle (x)$. 
In addition to the time dependent mass term due to 
$\langle \cx \ca^2\rangle (x)$ in decay mode $b$, the time evolution of 
expectation values in the two decay modes $a$ and $b$ are different, and 
therefore the condensation evolution for these modes are different too. 
Nonetheless there can be an exception to this argument. If we consider $A$ as a single quantum field and not
a collective notation, it is conceivable that at very early time symmetries
keep it massless and it receives a dynamical mass after an epoch of symmetry 
breaking and phase transition. In this case the classical field $\chi$ will
have a faster gross and at later times, a sudden change in the time evolution 
of the expectation values (\ref{valxaacal})-(\ref{valxaphical}) and $\chi$ is 
expected. Note that this argument is based on the back-reaction of the $X$ 
particles decay on $|\Psi\rangle$.

In the next subsection we show 
that the most important term in the late time evolution of $\chi$ is the 
non-homogeneous term in (\ref{evola}) and (\ref{evolb}). The coupling $g$ 
appears as a time independent constant and changes the amplitude of 
$\chi$. This means that the density of the condensate is proportional to 
$g^2 \propto \Gamma$. This confirms the results of the classical treatment of 
this model in Ref.~\cite{houridmquin}. Regarding the coincidence and smallness 
problem of the dark energy, the close relations between mass, number density 
and lifetime of $X$ and the mass and amplitude of the condensate is an 
evidence for an intrinsic feedback between dark matter and dark energy in 
this model. Moreover, this shows that the density of the dark energy is not 
more fine-tuned than the mass difference between left and right neutrinos in 
seesaw mechanism. In fact this similarity hints to a seesaw-like mechanism 
for the large difference between the mass of $X$ and $\Phi$ and supports 
the idea of right neutrino/sneutrino for $X$. 

Although here we are mostly interested in the decay of a meta-stable heavy 
particle, it is also interesting to see how a light scalar produced during 
the fast decay of another field, e.g. inflaton, can condensate and whether 
this condensate can last for long time and contribute to the build up of the 
dark energy. As before, in this setup the expectation values 
(\ref{valxaacal})-(\ref{valxaphical}) will have an initial rise. But, 
depending on the lifetime of $X$ particles is shorter or longer than the 
turning point time scale, either the latter happens earlier and $\chi$ rapidly
changes its slope from power-law to exponential decrease soon after the 
turning 
point, or power-low behaviour lasts for much longer time and the exponential 
break happens late in the lifetime of $\chi$. The former case is equivalent to 
having no additional terms due to interactions in 
(\ref{evola}) and (\ref{evolb}), and therefore after an initial build up, the 
condensate must fade quickly and will have no or very small contribution in
the dark energy. In the latter case the contribution can be yet significant.

In the approximation explained here, $k$-dependent part of the integral 
(\ref{appint}) is factorized and is proportional to $F(k)$ defined in 
(\ref{appk}). It depends on the dimensionless variable:
\be
K \equiv \frac{k^2}{B} = \frac {{k^2/a^2_0}}{H_0 m} \label{dimlessk}
\ee
The contribution of fields having $|\beta z_0| < 1$ grows exponentially as $K$
becomes larger- most probably due to short range quantum fluctuations of the 
expectation values, but this behaviour is transient and eventually arrives to 
saturation. On the other hand, for fields with $|\beta z_0| \gg 1$, 
$F (K) \propto (c + K^2) ^{-1}$. For large $K$ (small distances) $F(k)$ 
decreases with an slope index of $-2$, and for small $K$ (large distances) it 
approaches a constant. At late times all the fields are in this regime and 
therefore this presents the late time behaviour of the expectation values. 
Therefore $k$-independence is consistent with the observations of the dark 
energy at large scales.

Finally, we note that the time dependence of (\ref{appint}) is determined by 
$m/H_0$. This is similar to decoherence condition for massive scalar fields 
during inflation~\cite{decohmass}. Only fields with small $m/H_0$ can decohere 
and make classical density perturbations. This is another evidence for close
relation between these processes as suggested in Appendix \ref{app:a}.

\subsection{Evolution of the condensate} \label{sec:appcondens}
Now we use the result of the previous sections to estimate the 
evolution of the classical field $\chi$ with cosmic time. Neglecting the 
self-interaction term in (\ref{evola}) and (\ref{evolb}), the 
general solutions of these equations at the WKB approximation level are:
\bea
\chi (\eta) &=& \chi_1 D_q (\alpha_\Phi \eta) + \chi_2 D_q (-\alpha_\Phi \eta) + 
g \int_{\eta_0}^\eta d\eta'  \langle \cx \ca^2\rangle G^\chi (\eta,\eta')
\quad \quad \text{For decay mode $(a)$} \label {gensola} \\
\chi (\eta) &=& \chi^{'}_1 D_q \biggl (\alpha_\Phi \int d\eta (1 - 
\frac {2g \langle \cx \ca\rangle}{a^2 m^2_\Phi})^{\frac{1}{4}}\biggr ) + 
\chi^{'}_2 D_q \biggl (-\alpha_\Phi \int d\eta (1 - 
\frac {2g \langle \cx \ca\rangle}{a^2 m^2_\Phi})^{\frac{1}{4}} \biggr ) + 
\nonumber \\
&& 2g \int_{\eta_0}^\eta d\eta' \langle \Upsilon \cx \ca\rangle 
G^\chi (\eta,\eta') \quad \quad \text{For decay mode $(b)$} \label {gensolb}
\eea
where $\chi_1$, $\chi_2$, $\chi^{'}_1$ and $\chi^{'}_2$ are integration 
constants. We assume the following physically motivated initial conditions for 
these solutions:
\be
\chi (\eta_0) = \chi'(\eta_0) = 0 \label{chisolinit}
\ee
The associated homogeneous equation of the differential equation (\ref{evola}) 
is the same as equation (\ref{eqhomo}). Therefore the Green's function 
$G^\chi (\eta,\eta')$ is the 
same as the Feynman propagator (\ref{progexpand}). In the case of mode (b) we 
have added the WKB approximation to the homogeneous solutions. This correction 
is first order in the coupling $g$, thus we can use the Green's function
(\ref{progexpand}) for this mode too. A more precise solution can be obtained 
by replacing $D_q (\alpha_\Phi \eta)$ terms in (\ref{progexpand}) with the WKB 
corrected homogeneous solution $D_q (-\alpha_\Phi \int d\eta (1 - 
\frac {2g \langle \cx \ca\rangle}{a^2 m^2_\Phi})^{\frac{1}{4}})$.

Using the asymptotic behaviour of $D_q (z)$ at large $z$, we find that for 
both decay modes (a) and (b) the homogeneous part of the solutions 
(\ref{gensola}) and (\ref{gensolb}) is proportional to $\eta^{-1/2}$ and 
therefore deceases with time.
In the same way we can find time dependence of the special solution. For 
large $\eta$ it is proportional to $\eta ^{2+\epsilon}$ for both decay modes. 
Here $\epsilon$ is added by hand to parametrize the unknown time variation of 
the cosmological state $|\Psi\rangle$. If the mass of $\Phi$ is very small
such that $B_{\Phi} \approx 0$, the solution of the evolution equations is
approximately a plane wave in $\eta$ coordinate and $\chi \propto \eta ^{3+
\epsilon}$ for the mode (b). For mode (a) this approximation does not modify 
the asymptotic behaviour of $\chi$.

Finally the classical field $\varphi$ in comoving coordinates is 
$\chi = a \varphi$ and varies as:
\bea
\varphi \propto t ^{-1/2+2+\epsilon} && \text{Assuming $m \neq 0$ for all 
fields, both decay modes.} \label{phiasympm} \\
\varphi \propto t ^{-1/2+3+\epsilon} && \text{For $m_{\Phi} \approx 0$ and 
mode (b).} \label{phiasympb}
\eea
For all cases, the late time behaviour of $\varphi$ depends on $\epsilon$
i.e. the back reaction. Based on (\ref{abpsi}) and qualitative arguments, we 
expect that $\epsilon$ depends on the density and lifetime of $X$ field, and 
therefore on the expansion of the Universe. If we want that at late times 
$\mathbf {\varphi \sim cte.}$, the only consistent model for the quintessence 
field is a model in which $\mathbf {m_{\Phi} \neq 0}$ but small. In such a 
model the condensate grows during the radiation dominated era but stops 
growing when matter becomes dominant. However, self-interaction should 
somehow modify this result. None the less, the strong limit on the clustering 
of the dark energy shows that self-interaction of the quintessence field can 
not be very strong and therefore its over all effect must be small. This 
simple model considered here favors particle physics models with a PNGB 
scalar field and a cyclical potential to protect the small mass of 
$\mathbf{\Phi}$~\cite{png}. 

At late times when the 
energy density of the dark matter $X$ becomes comparable to the density of 
the dark energy $\sim 1/2 m_{\Phi}^2 \varphi^2$, there is a 
strong feedback between expansion rate and the density of dark matter. Faster 
expansion rates bring down the density of the dark matter and therefore the 
rate of $\varphi$ production decreases. This reduces the density of the dark 
energy and the expansion. We have seen the same feedback in classical 
treatment of this process in Ref.~\cite{houridmquin}. 

\section{Outline}\label{sec:conclu}
Although we have not yet observed any elementary scalar field, we believe they 
play important roles in the foundation of fundamental forces in the nature, in 
Standard Model and in all suggested extension of it. The detailed studies of 
their behaviour and their consequences for other phenomena are however 
hampered by the fact that they can have complex non-linear self-interaction 
and interaction with other fields which make analytical calculations very 
difficult. 

We used quantum field theory techniques to determine the evolution of the 
classical component - the condensate - of a scalar field produced during the 
decay of a much heavier particle. Such a process had necessarily happened 
during the reheating of the Universe, and similar phenomena can happen - both 
as decay and as interaction - after the reheating if a scalar has interaction 
with other fields. We showed that a significant amount of condensate can only 
be obtained for light fields with masses comparable or smaller than Hubble 
constant at the initial time. By considering two decay modes, we also showed 
that the type of interactions has very important role in the cosmological 
evolution of the condensate and its contribution to dark energy. Therefore, 
the details of such a model depends on the particle physics of the fields. 
The general behaviour is nonetheless universal. Due to the coupling between 
quantum phenomena, i.e the micro-physics of the Universe, and its classical 
macroscopic content, obtaining an exact analytical solution is impossible. 
We left this task for future and by using simplifying assumptions we deduced 
the general aspects of the asymptotic time evolution of the condensate. We 
showed that one of the most important ingredient of this model is the unknown 
back-reaction of the decay and other physical processes on the Fock space 
of the Universe. We parametrized the time dependence of the Fock space with 
just one exponent $\propto \eta^\epsilon$. If during matter dominated epoch 
$-4 \lesssim \epsilon \lesssim -2$, the condensate density evolves very slowly 
with time and has a behaviour similar to the observed dark energy. This range 
of $epsilon$ is for minimal model without self-interaction for the scalar 
field. In a realistic context, both self-interaction and interaction with 
other field, specially Standard Model fields must be considered and they can 
change this range.

The main goal of this work has been studying the contribution of the classical 
component of a gradually built-up field in the dark energy, specially in the 
context of a metastable heavy particle. Nonetheless, most of the 
results obtained here are applicable to other contexts and other models. 
Specifically, extensions of the Standard Model such as supersymmetric and 
supergravity models and string theory contain a large number of scalar fields. 
Condensation of these fields and their evolution and thereby their parameter
space and symmetries can be constrained by the observation of the equation 
of state of the Universe. As we have shown here, the condensate evolution
depends on a number of parameters including mass, decay rate of the producing 
particle, interactions and couplings. This is an additional and more detailed 
information with respect to simple density constraint for the scalar fields of 
particle physics models.

\vspace{1.5cm}

\begin {appendix}
{\Large \bf Appendixes}
\section{A note on the relation between decoherence and condensation} 
\label{app:a}
Using canonical 
representation and Bogolubov transformation for a free scalar field, one can 
find the relation between creation and annihilation operators at two different 
cosmic times:
\bea
\begin{pmatrix}a_k(\eta) \\ a^{\dagger}_k(\eta) \end{pmatrix} & = &
\begin{pmatrix} u_k (\eta, \eta_0) & v_k (\eta, \eta_0) \\ 
u^*_k (\eta, \eta_0) & v^*_k (\eta, \eta_0) 
\end{pmatrix}\begin{pmatrix}a_k({\eta}_0) \\ a^{\dagger}_k({\eta}_0) 
\end{pmatrix}\label {boglio} \\
\Phi (\eta) & = & \frac {-i a}{\sqrt {2 |k^2 + a^2 m^2 - 
\frac{a''}{a}|}} (a^{\dagger}_k(\eta) + a_k(\eta))\label {phiexp}
\eea
where $a^{\dagger}_k({\eta}_0)$ and $a_k({\eta}_0)$ are respectively creation 
and annihilation operator at an initial conformal time ${\eta}_0$, and 
$u_k (\eta, \eta_0)$, $v_k (\eta, \eta_0)$ and their conjugates are 
proportional to the solutions of the field equation at $\eta$ and 
$\eta_0$~\cite{decohmass}. Using the definition of number operator 
$\hat N \equiv a^{\dagger}_ka_k$, it is 
easy to show that even without any self-interaction the number of particles 
at $\eta$ and $\eta_0$ are not equal and cosmic expansion leads to particle 
production. The decoherence of these particles however needs an interaction to 
couple modes or fields~\cite{coarsegrain}~\cite{decohereint}. Although in a 
curved spacetime, specially during inflation, squeezed states can be 
achieved~\cite{squeezed}, the decoherence is not complete unless an 
interaction breaks the entanglement between degenerate quantum 
states by coupling causally unrelated modes - modes inside particle horizon to 
ones outside~\cite{infdecohere}~\cite{decohereint}. 

We observe that for a free field in which modes are independent, the linearity 
of (\ref{boglio}) and the properties of the creation/annihilation operators 
leads to $\langle 0|\Phi (\eta)| 0\rangle = 0$, where $| 0\rangle$ is the vacuum 
at $\eta_0$, i.e. there is no condensation. Therefore decoherence is a necessary 
but not sufficient condition for the formation of a condensate.

\section{Free field Green's function on non-vacuum states} \label{app:b}
In canonical representation, a free scalar field $\phi$ can be decomposed to 
creation and annihilation operators of an orthogonal basis of the Fock space:
\be
\phi (x) = \sum_k \um_k (x) a_k + \um_k^* (x) a^{\dagger}_k \quad , \quad 
[a_k, a^{\dagger}_{k'}] = \delta_{kk'} \quad [a_k,a_{k'}] = 0 \quad 
[a^{\dagger}_k,a^{\dagger}_{k'}] = 0\label{canon}
\ee
where $\um_k (x) \equiv \um_k (\eta)e^{-i\vec{k}\vec{x}} $ is a solution of 
the free field equation (\ref{eqhomo}). Quantization of $\phi$ imposes the 
following relation:
\be
\um_k (\eta,x)\um_k^{'*}(\eta,y) - \um_k^{'}(\eta,x) \um_k^*(\eta,y) = i 
\delta^{(3)} (x-y) \label{quantcond}
\ee
A Fock state $|\Psi\rangle$ is constructed by multiple applications of the 
creation operator $a^{\dagger}_k$ on the vacuum state $|0\rangle$ defined by:
\bea
&& a_k |0\rangle = 0,~ \forall k \quad , \quad |k_1 k_2 \ldots k_n\rangle 
\equiv a^{\dagger}_{k_1} a^{\dagger}_{k_2} \ldots a^{\dagger}_{k_n} |0\rangle 
\label{vacdef} \\
&& |\Psi\rangle = \sum_{k_1 k_2 \ldots k_n} \Psi_{k_1 k_2 \ldots k_n}
|k_1 k_2 \ldots k_n\rangle \label{psistate}
\eea
Applying these decompositions to 2-point free Green's function of $\phi$, it 
can be written as:
\bea
iG_F (x,y) &\equiv &\langle\Psi|T\phi(x)\phi(y)|\Psi\rangle = \nonumber \\
&& \sum_k \sum_i \sum_{k_1 k_2 \ldots k_n} \delta_{kk_i} |\Psi_{k_1 k_2 
\ldots k_n}|^2 \biggl [\um_k^* (x)\um_k (y) \Theta (x_0 - y_0) + 
\um_k (x)\um_k^* (y) \Theta (y_0 - x_0)\biggr ] + \nonumber \\
&& \sum_k  \biggl [\um_k (x)\um_k^* (y) \Theta (x_0 - y_0) + \um_k^* (x)
\um_k (y) \Theta (y_0 - x_0)\biggr ] \label{propst}
\eea
From (\ref{propst}) we can extract the expression for future and past propagators:
\bea
iG^> (x,y) &\equiv& \langle\Psi|\phi(x)\phi(y)|\Psi\rangle = \sum_k \sum_i 
\sum_{k_1 k_2 \ldots k_n} \delta_{kk_i} |\Psi_{k_1 k_2 \ldots k_n}|^2 
\um_k^* (x)\um_k (y) + \nonumber \\
&& \sum_k \biggl [1 + \sum_i \sum_{k_1 k_2 \ldots k_n} \delta_{kk_i} 
|\Psi_{k_1 k_2 \ldots k_n}|^2 \biggr ] \um_k (x)\um_k^* (y) \label{propfu} \\
iG^< (x,y) &\equiv& \langle\Psi|\phi(x)\phi(y)|\Psi\rangle = \sum_k \sum_i 
\sum_{k_1 k_2 
\ldots k_n} \delta_{kk_i} |\Psi_{k_1 k_2 \ldots k_n}|^2 \um_k (x)\um_k^* (y) + 
\nonumber \\
&& \sum_k \biggl [1 + \sum_i \sum_{k_1 k_2 \ldots k_n} \delta_{kk_i} 
|\Psi_{k_1 k_2 \ldots k_n}|^2 \biggr ] \um_k^* (x)\um_k (y) \label{proppa} 
\eea
Therefore, for free fields, $G_F (x,y)$ on any density operator $\rho$ can 
be written as a linear expansion with respect to $\um_k (x)\um_k^* (y)$ and 
$\um_k^* (x)\um_k (y)$, the independent solutions of the free field equation. 
The contribution from a non-vacuum 
state $|\Psi\rangle$ appears in the coefficients of the expansion, just as the 
initial/boundary condition effects appear in the expansion coefficients of
the propagator (\ref{progexpand}). Projection coefficients $|\Psi_{k_1 k_2 
\ldots k_n}|^2$ determine the momentum distribution of the states in the 
environment. In the simplest case where quantum correlation between particles 
is negligible, it is proportional to one particle momentum distribution $f(k)$ 
that in the case of a thermal environment is Boltzmann, Fermi or 
Bose-Einstein distribution.

Here a general comment about expectation values including 2-point Green's 
function is in order. An expectation value i.e. an inner product in the Fock 
space of a quantum system, this includes operators at different spacetime 
points is meaningful only if these points share the same or isomorphic Fock 
spaces. The state $|\Psi\rangle$ in (\ref{propst}) for which the expectation 
value is calculated must be a member of the Fock space at $x$ and $y$. 
However, in the fast changing early Universe, it is not evident that such a 
condition exists unless $x$ and $y$ are enough close to each other. This 
reflects the back-reaction of interactions during and after reheating on the 
Fock space and coupling between operators and states as mentioned in 
Sec.\ref{sec:wkbprog}. Therefore, $|\Psi\rangle$ has an implicit 
spacetime dependence and one-particle distribution function gets the familiar 
form $f(x, k)$ used in the construction of classical Boltzmann 
equation~\cite{collterm}.

\section{Propagators and evolution in matter dominated epoch} \label{app:c}
In the matter dominated epoch the relation between comoving and conformal 
time is defined as:
\bea
 \eta = \int \frac{dt}{a} = \eta_0 \biggl( \frac{t}{t_0}\biggr )^{\frac{1}{3}} 
&,& \eta_0 \equiv \frac{3t_0}{a_0} \label{etamatter} \\
 \frac{a}{a_0} = \biggl( \frac{t}{t_0}\biggr )^{\frac{2}{3}} = 
\biggl( \frac{\eta}{\eta_0}\biggr )^2 &,& \frac{a''}{a} = \frac{2}{\eta^2} 
\label{amatter}
\eea
By applying (\ref{amatter}) to the field equations (\ref{evola}) and 
(\ref{evolb}) and neglecting interactions, the field equation for both modes 
is the following:
\be
\chi'' + (k^2 + \frac{m^2 a_0^2\eta^4}{\eta_0^4} - \frac{2}{\eta^2}) \chi = 0 
\label{evolmatter}
\ee
where $\chi$ presents one of the fields $\chi$, $\Upsilon$, $\cx$ or $\ca$. 
For two special cases of $m = 0$ and $k^2 = 0$ this equation has exact 
analytical solutions:
\be
\chi (\eta) = \begin{cases} \sqrt{\eta} J_{\pm \frac{1}{2}} (\beta \eta^3) 
\quad , \quad \beta \equiv \frac{\sqrt{3} m a_0}{3\eta_0^2} & 
\text{For $k^2 = 0$} \\
\sqrt{\eta} J_{\pm \frac{3}{2}} (k \eta) & \text{For $m = 0$}
\end{cases} \label {solkmmatter}
\ee
As usual, in each case the WKB approximation can be used to find an 
approximate solution when the neglected terms are not zero. Another possible 
approximation is a linear interpolation between two cases in 
(\ref{solkmmatter}). 

Bessel functions $J_{\pm 1/2}$ and $J_{\pm 3/2}$ have close analytical 
expressions:
\bea
J_{\frac{1}{2}} (x) = \sqrt{\frac{2}{\pi x}} \sin x &,&
J_{-\frac{1}{2}} (x) = \sqrt{\frac{2}{\pi x}} \cos x \label {besselonehalf} \\
J_{\frac{3}{2}} (x) = \frac {1}{x} J_{\frac{1}{2}} (x) - J_{-\frac{1}{2}} (x) 
&,& J_{-\frac{3}{2}} (x) = -J_{\frac{1}{2}} (x) - \frac {1}{x} 
J_{-\frac{1}{2}} (x) \label {besselthreehn}
\eea
and 
\be
\chi (\eta) = \begin{cases} \frac{\sqrt{\frac{2}{\pi \beta}}}{\eta} \biggl
 (\chi_1 \sin(\beta \eta^3) + \chi_2 \cos(\beta \eta^3)\biggr ) & 
\text{For $k^2 = 0$} \\
\sqrt{\frac{2}{\pi k}} \biggl (\chi'_1 (\frac{\sin (k\eta)}{k\eta} - 
\cos (k\eta)) + \chi'_2 (-\sin (k\eta) - \frac{\cos (k\eta)}{k\eta})\biggr ) & 
\text{For $m = 0$} \end{cases} \label {solkmmatterexp}
\ee
where $\chi_1$, $\chi_2$, $\chi'_1$ and $\chi'_2$ are integration constants. 
At late times and for large scales (small $k$) the mass term in 
(\ref{evolmatter}) is dominant and $k^2 = 0$ approximation can be applied. 
Using (\ref{solkmmatterexp}), we find that integrals analog to 
(\ref{prointeg}) (replacing $D_q$ with $\chi$ from (\ref{solkmmatter})) have 
a late time behaviour $\propto \eta^{9-2N}$ where $N$ is the number of fields 
in the expectation value. If we assume that the mass of $\Phi$ is very small 
and $m = 0$, this approximation can be applied to this field. Under these 
conditions the expectation values containing one $\chi$ field are 
$\propto \eta^{9-2(N-1)}$ where here $N$ is the number of other fields. 
Finally from these results and the general solutions (\ref{gensola}) and 
(\ref{gensolb}) (again replacing ${\mathcal U}_k$ with $\chi$) we conclude 
that:
\bea
\varphi \propto t^{\frac{2+\epsilon}{3}} &,& \text{For $m \neq 0$, both modes} 
\label{phimattera} \\
\varphi \propto t^{\frac{4+\epsilon}{3}} &,& \text{For $m_{\Phi} = 0$ and 
mode (b)} \label{phimatterb}
\eea
Comparing (\ref{phimattera}) and (\ref{phimatterb}) with (\ref{phiasympm}), 
and (\ref{phiasympb}) and assuming the same $\epsilon$, we find that in the 
matter dominated epoch the time evolution is slower than radiation dominated 
epoch. In the matter dominated epoch the expansion of the Universe is faster,
the decay of $X$ particles is slower because for the same average density the 
Universe is younger, and therefore less $X$ particles have decayed. This 
compensates the density reduction due to a faster expansion. Similar to the  
radiation dominated epoch, the final time evolution rate depends on the 
unknown parameter $\epsilon$.
\end{appendix}

\end{fmffile}

\end{document}